\documentclass[aps,onecolumn,showpacs,amsmath,amssymb,pra,preprint]{revtex4-1}

\usepackage{graphicx}%
\usepackage{dcolumn}
\usepackage{bm}
\usepackage{braket}
\begin{document}

\title{Time-resolved  dynamical Franz-Keldysh effect under elliptically polarized laser}
\author{T. Otobe}
\affiliation{Kansai Photon Science Institute, National Institutes for Quantum and Radiological Science and Technology (QST), Kyoto 619-0215, Japan}
\begin{abstract}
The analytical formula for the time-resolved dynamical Franz-Keldysh effect (Tr-DFKE) under an elliptically polarized laser in sub-femtosecond time-scale is reported. 
The Houston function is assumed as the time-dependent wave function of the parabolic two-band system. 
The resulting formula shows the sub-cycle change of the optical properties for elliptically polarization; the modulation of the dielectric function becomes smaller than that of linear polarization. 
On the other hand, the subcycle modulation of the dielectric function disappears for a circularly polarized laser, 
which is a significant feature of the Tr-DFKE.
This analytical formulas show good qualitative agreement with the first-principle calculation employing the time-dependent density functional theory for diamond.
\end{abstract}
\maketitle
\section{Introduction}
In the last two decades, advances in laser sciences and technologies have led to the availability of intense coherent light with different characteristics.
 Ultra-short laser pulses can be as short as a few tens of an attoseconds, leading to the new field of attosecond science
\cite{atto01}.  
Intense laser pulses of mid-infrared (MIR) or terahertz (THz) frequencies  
have also recently become available \cite{HDBT11, Chin01}.  
By employing these extreme sources of coherent light, 
it is possible to investigate the optical response of materials in real time with a resolution much lower than an optical cycle 
\cite{atto01,Hirori11,Krausz13,Schultze13,Schultze14,Novelli13}.  
 
The dielectric function $\varepsilon(\omega)$ is the most fundamental  
quantity characterizing the optical properties of matter.  
The dielectric function observed in an ultra-fast  pump-probe experiment
can be  further considered as a  probe time ($T_p$)-dependent function, $\varepsilon(T_p,\omega)$.
The modulation of the dielectric function  $\varepsilon(\omega)$  in the presence of  
electromagnetic fields has been a subject of investigation  
for many years. The change in the dielectric function under a static electric field is known  
as the Franz-Keldysh effect (FKE)   
\cite{Franz58,Keldysh58,Tharmalingam63,Seraphin65,Nahory68,
Shen95,Sipe10,Sipe15}, 
and the change under an alternating electric field is known as  
the dynamical FKE (DFKE)  
\cite{Yacoby68, Jauho96, Nordstorm98, Ajit04,Mizumoto06, Shambhu11}. 

We determined the sub-cycle change in the optical properties, i.e., the time-resolved DFKE (Tr-DFKE),
which corresponds to the response of the dressed states and quantum path interference of different dressed state \cite{otobe16}.
In particular, this ultra-fast change exhibits an interesting phase shift that depends on the field amplitude and probe frequency. 
By utilizing this phenomenon, we can develop an ultra-fast modulator of light or an ultra-fast optical switch.
Recently, the Tr-DFKE was observed experimentally by a near-infrared (NIR)-pump  extreme ultraviolet (EUV)-probe with attosecond time-resolution for polycrystalline diamond \cite{Lucchini16}.
A similar effect was also observed in an excitonic state in a quantum well of GaAs by a THz-pump NIR-probe \cite{Uchida16}.

We  previously showed that the field amplitude of the pump laser and the bandwidth of the probe pulse are crucial parameters for controlling the Tr-DFKE \cite{otobe16}.
Another possible control parameter is the polarization of the laser.
In the current work, we present  analytical formulas and  first-principle calculations for the Tr-DFKE under elliptically  and circularly polarized pump lasers.
The time-dependent oscillation of the optical properties decreases as the ellipticity increases.
We found that the time-dependent change in the optical properties completely disappears under a circularly polarized pump laser for an isotropic band structure.
We compare our formulas with first-principle calculations employing time-dependent density-functional theory (TDDFT) for  diamond.

 The remainder of this paper is organized  as follows.
 In Sec. II, we  develop an analytical formulation for the Tr- DFKE under a circularly polarized light 
 using a parabolic two-band model.
 In Sec. III, we present a real-time TDDFT approach for the Tr-DFKE.
 In Sec. IV, we present the numerical results.
 In Sec. V, a summary is given.
\section{Formulation}
To derive the time-dependent conductivity, we will revisit a simple model that we reported in our previous work \cite{otobe16}.
The probe's electric field is assumed to be weak enough that is can  be treated  
using the linear response theory. We denote the electric current caused by  
the probe field as $J_p(t)$, which is assumed to be parallel to the
direction of the probe's electric field. 
the relationship to  the time-domain  
conductivity $\sigma(t,t')$ is given as:
\begin{equation} 
J_p(t) = \int_{-\infty}^t dt' \sigma(t,t') E_p(t'), 
\label{def_sigma} 
\end{equation} 
where $E_p(t')$ is the electric field of the probe pulse.
We note that the conductivity $\sigma(t,t')$ depends on both 
times $t$ and $t'$ rather than the just the time difference
$t-t'$, owing to the presence of the pump pulse.
If the probe laser is much shorter than  the optical cycle of the pump laser and 
has peaks at time $t=T_p$, we can define the time-dependent conductivity, $\tilde{\sigma}(T_p,\omega)$, as:
\begin{equation}
\tilde{\sigma}(T_p, \omega) =\frac{\int dt e^{i\omega t} J_p(t)}{\int dt E_p(t)}.
\end{equation}

In the following, we consider a simplified description:
electron dynamics in the presence of the pump and probe fields are 
assumed to be described by a time-dependent Schr\"odinger 
equation for a single electron:  
\begin{equation}
 i\frac{\partial}{\partial t} \psi_n(\vec r,t) 
= \left[\frac{1}{2m_e}\left(\vec p+\frac{e}{c}\vec A(t)\right)^2+V(\vec r) \right] \psi_n(\vec r,t), 
\label{TDSE}
\end{equation}
where $\psi_n(\vec r,t) $ is the time-dependent wave function of the $n$-th band, 
 $\vec A(t)$ is the vector potential of the pump light field, and 
 $V(\vec r)$ is a time-independent, lattice periodic potential.
 In this paper, we employ atomic units for all equations.
We express the solution of this equation using the time-dependent 
Bloch function $v_{n\vec k}(\vec r,t)$ as
$\psi_n(\vec r,t)=\sum_{\vec{k}}e^{i\vec k \vec r} v_{n \vec k}(\vec r,t)$, 
where $\vec{k}$ is the  Bloch wavenumber vector.  

We further assume that in the presence of the pump field
described by a vector potential $\vec A_P(t)$,  the solution of 
Eq. (\ref{TDSE}) is well approximated by the  Houston 
function \cite{Yacoby68,Houston}. 
Using static Bloch orbitals  $u_{n\vec k}(\vec r)$ and orbital energies 
$\epsilon_{n\vec k}$  that satisfy:
\begin{equation}
\left[\frac{1}{2m_e}\left(\vec p+\vec k \right)^2+V(\vec r) \right] u_{n\vec k}(\vec r)
= \epsilon_{n\vec k} u_{n\vec k}(\vec r),
\end{equation}
the Houston function can be expressed as:
\begin{equation} 
 w_{n\vec{k}}(\vec{r},t)=u_{n \, \vec k_P(t)}(\vec{r}) 
 \exp \left[-i \int^t \epsilon_{n \,\vec k_P(t')} dt' \right],
 \end{equation} 
where $\vec k_P(t)$ is defined by $\vec k_P(t) = \vec k + e\vec A_P(t)/c$.

We now consider an elliptically polarized pump laser with the vector potential:
\begin{equation}
\vec{A}_P(t)=A_0(\eta \sin\Omega t,0,\cos \Omega t)~~ (0\leq \eta \leq 1),
\end{equation}
and a two-band model, considering of
only the two orbitals in the  occupied valence ($v$) and unoccupied conduction
($c$) bands. 
\textit{Here, $\eta$ is the ellipticity of the pump laser.}
The excitation energy from the valence band
to the conduction band is assumed to have a parabolic form:
\begin{equation}
\epsilon_{c \vec k} - \epsilon_{v \vec k}
\simeq \frac{k^2}{2\mu} + \epsilon_g\equiv \epsilon_k+\epsilon_g,
\end{equation}
where $\epsilon_g$ is the band gap energy and $\mu$ is the reduced mass
of the electron-hole pairs.
We assume the probe is a $\delta$-functional linear polarized laser, $f_p\delta(t-T_p)$.
Our formulas do not depend on the polarization of probe laser.

Using the same procedure as in our previous work \cite{otobe16}, we obtain the time-resolved 
conductivity:
\begin{eqnarray}
\label{E_DFKE}
&&\varepsilon_E(T_p,\omega)=1-\frac{4\pi e^2}{m\omega^2}n_e-\frac{2 e^2|p_{cv}|^2\mu^{3/2}}{\sqrt{2}m_e^2\pi} \int_0^{\infty} \sqrt{\varepsilon_k}d\varepsilon_k \int _{-1}^{1} d\cos\theta \int_0^{2\pi} d\phi\nonumber\\
&\times& \sum_{l_1,l_2,m,n} (-1)^{m-n}\tilde{J}_{l_1}(\alpha,\beta)\tilde{J}_{l_1-2m}(\alpha,\beta)J_{l_2}(\gamma)J_{l_2+2(m-n)}(\gamma)\nonumber\\
&\times&\left[ \frac{e^{i2n\Omega T_p}}{(\omega+2n\Omega)(\omega-(\varepsilon_g+\varepsilon_k+U_E +(l_1+l_2-2n)\Omega))} \right.\nonumber\\
&-&\left. \frac{e^{-i2n\Omega T_p}}{(\omega-2n\Omega)(\omega+(\varepsilon_g+\varepsilon_k+U_E +(l_1+l_2-2n)\Omega))}\right],
\end{eqnarray}
where:
\begin{eqnarray}
\alpha&=&\frac{e k A_0}{c\mu \Omega}\cos\theta, \\
\beta&=&\frac{e^2 A_0^2}{8c^2\mu \Omega}(1-\eta^2), \\
\gamma&=&\eta \frac{e k A_0}{c\mu \Omega}\sin\theta\cos\phi, 
\end{eqnarray}
and: 
\begin{equation}
U_e = \frac{e^2 A_0^2}{4c^2\mu}(1+\eta^2).
\end{equation}
Here $U_e$ is the average kinetic energy in the elliptic polarized laser field.
Here, $J_l$ is the $l$th Bessel function and
$\tilde{J}_l$ is the generalized Bessel function \cite{Reiss03} defined as:
\begin{equation}
 \tilde{J}_l(\alpha,\beta)=\sum_i J_{l-2i}(\alpha)J_i(\beta).
\end{equation}
We define the $\vec{k}$ by the polar angles, such that
$\vec{k}=k(\sin\theta\cos\phi, \sin\theta\sin\phi, \cos\theta)$.

Equation~(\ref{E_DFKE}) is connected to the Tr-DFKE   under a linearly polarized laser at $\eta=0$, which we reported in our previous paper \cite{otobe16}:
\begin{eqnarray}
\label{L_DFKE}
&&\varepsilon_L(T_p,\omega)= 1-\frac{4\pi e^2}{m_e\omega^2}n_e-\frac{ 4 e^2|p_{cv}|^2\mu^{3/2}}{\sqrt{2}\pi m_e^2\omega}
\int_0^{\infty} d\epsilon_k \sqrt{\epsilon_k} \int_{-1}^1 d(\cos\theta)  \sum_{l,n} \tilde{J}_l(\alpha,\beta)\tilde{J}_{l-2n}(\alpha,\beta) \nonumber\\
&\times& \left[ 
\frac{e^{i 2n\Omega T_p}}{(\omega+2n\Omega)\{\omega-(\epsilon_g+\epsilon_k+U_p +(l-2n)\Omega)\}} \right. \nonumber\\
&-&\left.  \frac{e^{-i 2n\Omega T_p}}{(\omega-2n\Omega)\{\omega+(\epsilon_g+\epsilon_k+U_p +(l-2n)\Omega)\}}\right],
\end{eqnarray}
where $U_p=\frac{e^2A_0^2}{4\mu c^2}$ is the well known ponderomotive energy.

The form of the circularly polarized laser ($\eta=1$) becomes:
\begin{eqnarray}
\label{C_DFKE}
&&\varepsilon_C(T_p,\omega)=1-\frac{4\pi e^2}{m_e\omega^2}n_e-\sum_{l}\frac{4 e^2|p_{cv}|^2\mu^{3/2}}{\sqrt{2}\pi m_e^2\omega^2}
\int_0^{\infty} \sqrt{\varepsilon_k}d\varepsilon_k\int _{-1}^{1} d\cos\theta J^2_{l}(\alpha_1) \nonumber\\
&\times&\left[ \frac{1}{\omega-(\varepsilon_g+\varepsilon_k+U_c+ l\Omega)} -\frac{1}{\omega+(\varepsilon_g+\varepsilon_k+U_c+ l\Omega)}\right],
\end{eqnarray}
where:
\begin{equation}
\alpha_1=\frac{ekA_0\sin\theta}{\mu c\Omega}
\end{equation}
 and: 
\begin{equation}
 U_c=\frac{e^2A_0^2}{2\mu c^2}.
\end{equation}
In this case, we redefine the polarization as:
\begin{equation}
\vec{A}_P(t)=A_0(\cos \Omega t, \sin\Omega t, 0),
\end{equation}
to have a simpler formula.
An interesting point is that the time-dependence of Tr-DFKE disappears for a circularly polarized laser since the summation over $\vec{k}$ corresponds to the time-average for one $\vec{k}$-point isotropic system.
We note that $\varepsilon_C$ does not depend on the parameter $\beta$ that corresponds to the parameter $\gamma=U_p/\Omega$ for the linear polarization \cite{Nordstorm98}.
\textit{Above formulas do not depend on the polarization direction of the probe laser since the isotropic band structure is assumed.}

\section{Real-time TDDFT}
In  the above section, we derived the analytical forms for Tr-DFKE that can be used  for real materials.
Phase-sensitive modulation attributed to the Tr-DFKE under a linearly polarized laser has been reported by Lucchuni {\it et al. }\cite{Lucchini16} which agree with our previous work \cite{otobe16}.
This result indicates that real-time calculation based on the TDDFT is a the reliable approach.

In this section, we revisit our first-principle approach. 
The detail of the computational methods have been already reported elsewhere \cite{Bertsch00,Otobe08,otobe16}.
In real-time TDDFT, we describe electron dynamics in a unit  
cell of a crystalline solid under a spatially uniform electric field $E(t)$. 
Treating the field by a vector potential $\vec A(t)=-c\int^t dt' \vec E(t')$, 
the electron dynamics are described by the 
time-dependent Kohn-Sham (TDKS) equation \cite{Runge84}: 
\begin{equation} 
i \frac{\partial}{\partial t} \psi_i(\vec{r},t)= 
\left[ \frac{1}{2m_e} \left( \vec p + \frac{e}{c} \vec A(t) \right)^2 
+ V(\vec r,t) \right] \psi_i (\vec r,t), 
\label{TDKS} 
\end{equation} 
where $V(\vec r,t)$ is composed of 
electron-ion,  the Hartree, and exchange-correlation potentials.  
We use a norm-conserving pseudopotential for the electron-ion  potential \cite{TM91,Kleinman82}. 
For the exchange-correlation potential, we employ an adiabatic  
local density approximation (LDA)\cite{PZ81}.  

We calculate electron dynamics in diamond, using a cubic unit cell  
containing eight carbon atoms. 
The TDKS equation is solved in  
real time and real space. 
A total of $20^3$ real-space grids are used  
for the unit cell, and $32^3$ grids are used for the $k$-points.  
The Taylor expansion method is used for the time evolution \cite{Yabana96}  
with a time step of $\Delta t = 0.02$ in atomic unit.  
The number of time steps is typically 75,000. 
An important output of the calculation is the
average electric current density as a function of time, $J(t)$, given by: 
\begin{equation} 
J(t) = -\frac{e}{m_eV}  \int_{V} d\vec r \sum_i 
{\rm Re} \psi_i^* \left( \vec p + \frac{e}{c}\vec A(t) \right) \psi_i + J_{NL}(t), 
 \label{current} 
\end{equation} 
where $V$ is the volume of the unit cell and  
$J_{NL}(t)$ is the current caused by non-locality of the pseudopotential. 

In practice, we use the following electric fields. 
The pump field is of the form:
\begin{equation}
\vec{E}_P(t) = E_{0,P} f_P(t) (\eta\cos \Omega t, 0, \sin\Omega t) , 
\end{equation}
with a central angular frequency of $\Omega$ set to $\Omega=0.4$ eV.
The field is turned on adiabatically by $f_P(t)$, given
by $f_P(t)=\sin^2\left( \frac{\pi }{2 T_P}t\right)$ for $0<t<T_P$  
and $f_P(t)=1$ for $t \ge T_P$ and $T_P=$10 fs. 
 
The probe field is of the form:
\begin{equation}
\vec{E}_{p}(t) = \hat{y}E_{0,p}  \sin(\omega_p t) 
\exp\left(-(t-T_p)^2/\zeta^2\right).
\end{equation}
To improve the numerical accuracy, we assume a linear polarized probe pulse laser that is vertical to the pump laser field.
The average frequency $\omega_p$ is set to 5.5 eV, 
which is equal to the calculated band gap energy  
of diamond in LDA. The field strength is set to 
$E_{0,p}=2.7 \times 10^{-3}$ MV/cm, which is small enough
to probe the linear response of the medium.  
The pulse
duration $\zeta$ is set to $\zeta= 0.7$ fs.  

The frequency-dependent conductivity is calculated as the ratio of the
Fourier-transformed current $J_p(t)$ and field $E_p(t)$,  given by:
\begin{equation} 
\sigma(T_p, \omega) =  
\frac{\int dt e^{i\omega t} g(t-T_p) J_p(t)}{\int dt e^{i\omega t} E_p(t)}, 
\label{conductivity} 
\end{equation} 
where $g(t)$ is a filter function 
to damp the current and suppress spurious oscillations arising
from the numerical cutoff in the integration. 
The time-dependent conductivity  is converted to the dielectric function using the formula:
\begin{equation}
\epsilon(T_p,\omega)=1+i \frac{4\pi}{\omega}\sigma(T_p,\omega).
\end{equation}
\section{Numerical results}

\begin{figure}
\includegraphics[width=90mm]{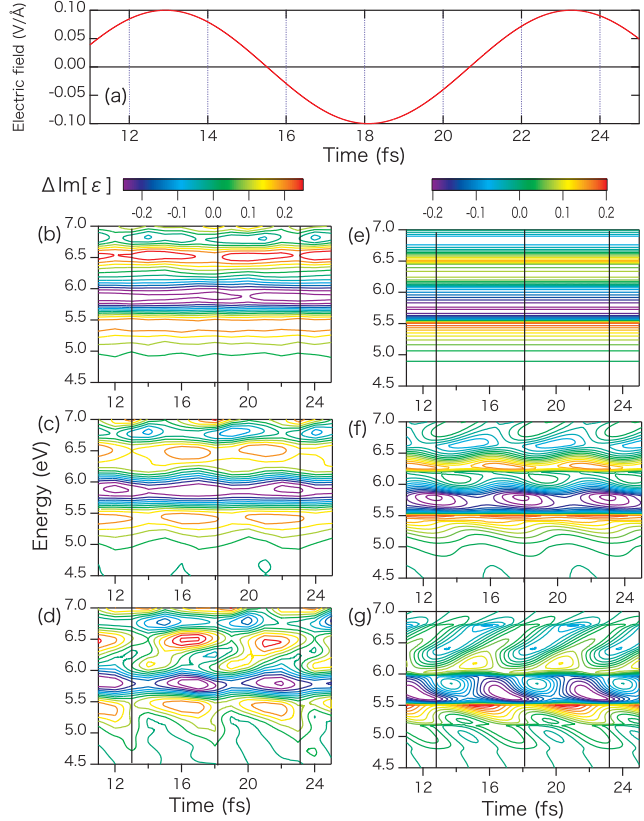}
\caption{\label{fig:Fig1}  Changes in $ {\rm Im}[\varepsilon(T_p,\omega)]$ ( $\Delta {\rm Im}[\varepsilon(T_p,\omega)]$) under circularly ((b) and (e)),  elliptically ((c) and (f)), and  linearly ((d) and (g)) polarized lasers.
(b)-(d) show the numerical results from the real-time TDDFT, and (e)-(g) show the results from analytical formulas  
 Eqs.~(\ref{E_DFKE}), (\ref{L_DFKE}), and (\ref{C_DFKE}), respectively. 
  Panel (a) shows the electric field of the pump laser. The horizontal axis presents the probe time, whose range  corresponds to the simulation by the real-time TDDFT.}
\end{figure}

\begin{figure}
\includegraphics[width=80mm]{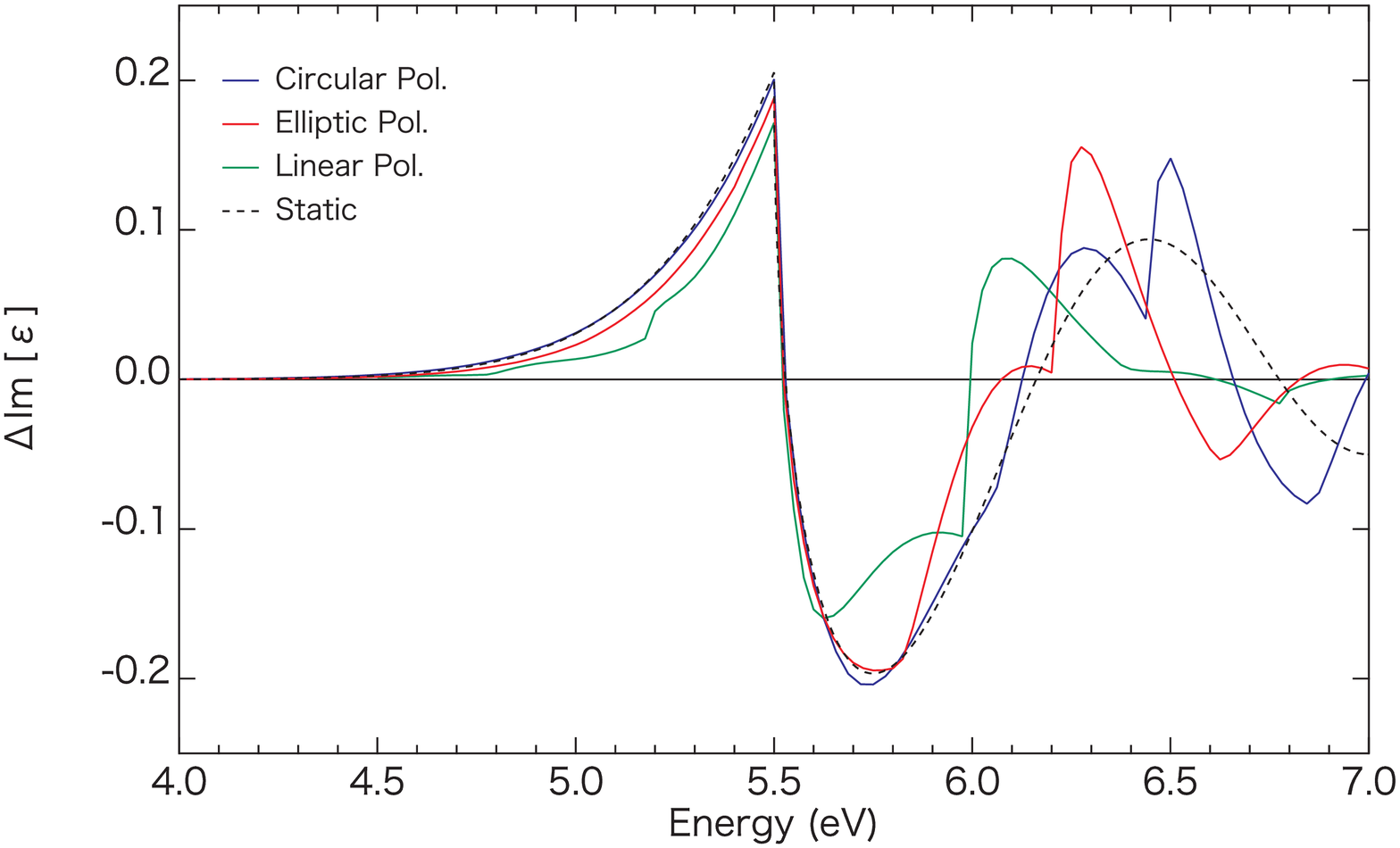}
\caption{\label{fig:Fig2} Time-average of the  $\Delta {\rm Im}[\varepsilon(T_p,\omega)]$ under a laser field with frequency of 0.4 eV and a field amplitude 10 MV/cm calculated from Eqs.~(\ref{E_DFKE}), (\ref{L_DFKE}), and (\ref{C_DFKE}). For reference, the static FKE calculated by the Tharmalingam formula \cite{Tharmalingam63} is represented by the black dashed line. }
\end{figure}

In this section, we aim to demonstrate the quantitative difference between circular and linear polarization 
by comparing the numerical results.
Diamond is a typical dielectric used in non-linear laser-matter interaction studies, and we   
selected it as an example with which to illustrate the application of the above formalism.

Figure \ref{fig:Fig1} shows the change in the imaginary part of the dielectric function, $\Delta {\rm Im}[\varepsilon(T_p,\omega)]$.
The maximum field amplitude is 10 MV/cm and the $\Omega$ is 0.4 eV.
Figure \ref{fig:Fig1} (b)-(d) show the results from the real-time TDDFT, and (e)-(g) show the results of the analytical theory. 
We assumed that the band gap, $\varepsilon_g$, is 5.5 eV,  the reduced mass $\mu$ is 0.25$m_e$, and the transition moment $P_{cv}=0.928$ in atomic units for the analytical formulas.
We adopt time components of $n=-1,0$, and 1 in Eqs.~(\ref{E_DFKE}), (\ref{L_DFKE}), and (\ref{C_DFKE}).
The vertical black solid lines in Fig.~\ref{fig:Fig1}(b)-(g)  represent the maximum of the absolute value of the electric field shown in Fig.~\ref{fig:Fig1} (a).
In numerical simulation by the real-time TDDFT, we assume the pump field rotating in $x$-$z$ plane, which corresponds to $a$- and $c$-axises of diamond, and the polarization of the probe pulse is set to be parallel to $b$-axis. 

The Tr-DFKE under circular polarization is presented in Fig.~\ref{fig:Fig1} (b) and (e) respectively.
As we expected in Eq.~(\ref{C_DFKE}), the time-dependence of DFKE almost disappears in real-time TDDFT.
However, for higher energy region around 6.5 $-$ 7eV, small oscillation of $\Delta  {\rm Im}[\varepsilon(T_p,\omega)]$ can be seen. 
The ratio of the small oscillation between the peak and dip is approximately 10 \%.
There are two possible explanations for this modulation.
First, the artificial error in real-time simulation is unavoidable. 
The Tr-DFKE  requires high computational accuracy. 
Therefore, small errors in time evolution and the Fourier transformation affect  the result.
The second is  the anisotropy of the band structure, which changes the definition of the reduced mass $\mu \rightarrow \mu(k,\theta,\phi)$.  
Although the band structure of diamond is almost isotropic around the $\Gamma$-point, it has a four-fold symmetric structure far from the $\Gamma$-point and higher energy bands.
Since the DFKE depends on the reduced mass,  time-dependence in circular polarization should include the effect of the structure of the valence and conduction bands.

Figures \ref{fig:Fig1} (c) and (f) show the case of  the elliptically polarized laser.
The real-time TDDFT  shows $2\Omega$ oscillation for all energies.
The analytic formula Eq.~(\ref{E_DFKE}) well reproduces the real-time TDDFT qualitatively for not only the frequency of the oscillation of $\Delta {\rm In} [\varepsilon(T_p,\omega)]$, but also the phase shift depending on the probe photon energy.
The amount of oscillation and the phase shift are weaker than that of the linear polarization shown in Figs.~\ref{fig:Fig1} (d) and (g).
However, the peaks of the modulation shows the same phase shift under  linear polarization.

Figure \ref{fig:Fig2} shows the time-average of  $\Delta {\rm Im}[\varepsilon(T_p,\omega)]$ of each polarization calculated by the analytic formula.
The time-average corresponds to the probe by continuous-wave lasers, and depends only on the density of states \cite{Jauho96}. 

The blue line denoted the case of the circularly polarized laser.
The exponential tail below the band gap, which is a feature of the FKE, can be seen.
The oscillation above the band gap is due to the response of the different dressed states and the blue shift of the band gap by $U_c$.  
In this case,  $U_c$ is approximately 0.95 eV, which corresponds to the energy at which $\Delta {\rm Im}[\varepsilon(\omega)]$ has a sharp peak.
Since the response  of the  $l=-1$ dressed state is intense, the apparent blue shift of the band gap becomes 0.5 eV.

The red line denotes the case of the elliptic polarization ($\eta=1/\sqrt{2}$).
The most obvious difference is the magnitude of the blue shift.
The original band gap is blue-shifted by $U_e=0.71$ eV. 
Although the overall behavior is similar to that of circular polarization, the contribution of the $l_1+l_2=-1$ dressed band becomes weaker than that of the circular polarization.

The case of linearly polarized light is represented by the green line.
Here, $U_p$ is 0.5 eV which is consistent with the apparent blue shift of the band gap.
There is a small shoulder  at 5.1 eV. This energy corresponds to the $l_1+l_2=-3$ dressed state.

For reference, the numerical result for the static FKE \cite{Tharmalingam63} is represented by the black dashed line.
Since the applied field amplitude is stable with circular polarization, the tunneling effect is expected to be similar to the static FKE. 
As expected, the behavior under the band gap agrees with the case of circularly polarized light.
However, the oscillation above the band gap is different.
This result indicates that the FKE under circularly polarized light is not  equivalent to the static FKE,
 even though the amplitude of the electric field is static.
Thus, using the dressed states is indispensable.  
\section{Summary}
We presented an analytical  formulation for the time-resolved dynamical Franz-Keldysh effect under an elliptically  polarized laser.
This formula connects the cases of linear polarization and circular polarization though the ellipticity of the pump laser.
We found that the time-dependent change in the optical properties observed under linearly polarized light disappears as the ellipticity increases. 
Therefore,  linearly polarized light is suitable for  ultrafast control of the material response.
On the other hand, the response of the dressed state is still important for understanding the change in the optical properties caused by a circularly polarized light field.
\section*{Acknowledgement}
This work is supported by a JSPS KAKENHI (Grants No. 15H03674). 
Numerical calculations were performed on the supercomputer SGI ICE X at 
the Japan Atomic Energy Agency (JAEA).


\begin{thebibliography}{99}
\bibitem{atto01} M. Hentschel, R. Klenberger, Ch. Spielmann, G.A. Reider, N. Milosevic, T. Brabec, U. Heinzmann,
M. Drescher, and F. Krausz, Nature {\bf 414}, 509 (2001). 
\bibitem{HDBT11} H. Hirori, A. Doi, F. Blanchard, and K. Tanaka, Appli. Phys. Lett. {\bf 98}, 091106 (2011). 
\bibitem{Chin01} A.H. Chin, O.G. Calderon, and J. Kono, Phys.Rev. Lett. {\bf 86} , 3292 (2001). 
\bibitem{Hirori11} H.Hirori, K. Shinokita, M. Shirai, S. Tani, Y. Kadoya, and K. Tanaka, Nat. Comm. {\bf 2}, 594 (2011).
\bibitem{Krausz13} A. Schiffrin, T. Paasch-Colberg, N. Karpowicz, V. Apalkov, D. Gerster,
S. M\"uhlbrandt, M. Korbman, J. Reichert, M. Schultze, S. Holzner, J.V. Barth, R. Kienberger, R. Ernstorfer, V.S. Yakovlev,
M.I. Stockman, and F. Krausz, Nature {\bf 493}, 70 (2013).
\bibitem{Schultze13} M. Schultze, E.M. Botschafter, A. Sommer, S. Holzner, W. Schweinberger, M. Fless, M. Hofstetter,
R. Kienberger, V. Apalkov, V.S. Yakovlev, M.I. Stockman, F. Krausz, Nature {\bf 493}, 75 (2013).
\bibitem{Novelli13} F. Nobelli, D. Fausti, F. Giusti, F. Parmigiani, and M. Hoffmann, Scientific Reports {\bf 3} 1227 (2013). 
\bibitem{Schultze14} M. Schultze, K. Ramasesha, C.D. Pemmaraju, S.A. Sato, D. Whitmore, A. Gandman, J.S. Prell,
L.J. Borja, D. Prendergast, K. Yabana, D.M. Neumark, S.R. Leone, Science {\bf 346}, 1348 (2014).
\bibitem{Franz58} W.Franz, Z. Naturforsch. Teil A {\bf 13}, 484 (1958). 
\bibitem{Keldysh58} L. V. Keldysh, Sov. Phys. JETP {\bf 34}, 788 (1958). 
\bibitem{Tharmalingam63} K. Tharmalingam, Phys. Rev. {\bf130}, 2204 (1963). 
\bibitem{Seraphin65} B. O. Seraphin and R. B. Hess, Phys. Rev. Lett. {\bf 14}, 138 (1965). 
\bibitem{Nahory68} R. E. Nahory and J. L. Shay, Phys. Rev. Lett. {\bf 21}, 1569 (1968). 
\bibitem{Shen95} H.Shen, M. Dutta, J. Appl. Phys. {\bf 78}, 2151 (1995).
\bibitem{Sipe10} J.K. Wahlstrand, J.E. Sipe, Phys. Rev. B{\bf 82}, 075206 (2010).
\bibitem{Sipe15} F. Duque-Gomez, J.E. Sipe, J. Phys. Chem. Solids {\bf 76}, 138 (2015).
\bibitem{Yacoby68} Y. Yacoby, Phys. Rev. {\bf 169}, 610 (1968). 
\bibitem{Jauho96} A. P. Jauho and K. Johnsen, Phys.Rev. Lett. {\bf76}, 4576 (1996). 
\bibitem{Nordstorm98} K. B. Nordstrom, K. Johnsen, S. J. Allen, A. P. Jauho, B. Birnir, J. Kono, T. Noda, H. Akiyama, and H. Sakaki, 
Phys. Rev. Lett. {\bf 81}, 457 (1998).
\bibitem{Ajit04} A. Srivastava, R. Srivastava, J. Wang, and J. Kono, Phys.Rev.Lett. {\bf 93}, 157401 (2004). 
\bibitem{Mizumoto06} Y. Mizumoto, Y. Kayanuma, A. Srivastava, J. Kono, and A. H. Chin, Phys. Rev. B {\bf 74}, 045216 (2006). 
\bibitem{Shambhu11} S. Ghimire, A.D. DiChiara, E. Sistrunk, U.B. Szafruga, P. Agostini, L. F. DiMauro, and D. A. Reis, Phys. Rev. Lett. {\bf 107}, 167407 (2011). 
\bibitem{otobe16} T. Otobe, Y. Shinohara, S. A. Sato, and K. Yabana, Phys. Rev. B {\bf 93}, 045124 (2016).
\bibitem{Lucchini16} M. Lucchini, S. Sato, J. Herrmann, A. Ludwig, M. Volkov, L. Kasmi, Y. Shinohara, K. Yabana, L. Gallmann, and U. Keller,  Science {\bf 353}, 916 (2016).
\bibitem{Uchida16} K. Uchida, T. Otobe, T. Mochizuki, C. Kim, M. Yoshita, H. Akiyama, L. N. Pfeiffer, K. W. West, K. Tanaka, and H. Hirori, JSAP-OSA Joint Symposia 2015 Abstracts, (Optical Society of America, 2015), paper 15p\_2E\_10. Paper is submitted.
\bibitem{Houston} W. V.  Houston, Phys. Rev. {\bf 51}, 184 (1940).  
\bibitem{Reiss03} H. R. Reiss and V. P. Krainov, J. Phys. A: Math. Gen. {\bf 36}, 5575 (2003).

\bibitem{Runge84} E. Runge and E. K. U. Gross,  Phys.Rev. Lett. {\bf 52}, 997 (1984).
\bibitem{Bertsch00} G.F. Bertsch, J.-I. Iwata, A. Rubio, and K. Yabana, Phys. Rev. B {\bf62} , 7998 (2000). 
\bibitem{Otobe08}  
T. Otobe, M. Yamagiwa, J. -I. Iwata, K. Yabana, T. Nakatsukasa, and G. F. Bertsch,  
Phys. Rev. B{\bf77}, 165104 (2008). 

\bibitem{TM91} N. Troullier and J.L. Martins, Phys. Rev. B {\bf 43}, 1993 (1991). 
\bibitem{Kleinman82} 
L. Kleinman and D. M. Bylander, Phys. Rev. Lett.  {\bf 48}, 1425 (1982). 
\bibitem{PZ81} J. P. Perdew and A. Zunger, Phys. Rev. B  {\bf 23}, 5048 (1981). 
\bibitem{Yabana96} K. Yabana and G.F. Bertsch, Phys. Rev. B {\bf 54}, 4484 (1996). 
\bibitem{yabana12} K. Yabana, T. Sugiyama, Y. Shinohara, T. Otobe, and G. F. Bertsch, Phys. Rev. B  {\bf 85}, 045134 (2012). 


\end{thebibliography}
\end{document}